\documentclass[12pt,epsfig]{article}
\usepackage{epsfig}
\newcommand{\singlespace}{
     \renewcommand{\baselinestretch}{1}\large\normalsize}
\newcommand{\doublespace}{
     \renewcommand{\baselinestretch}{1.5}\large\normalsize}

\newcommand{\be}{\begin{equation}}
\newcommand{\ee}{\end{equation}}
\newcommand{\ba}{\begin{eqnarray}}
\newcommand{\ea}{\end{eqnarray}}

\newcommand{\ave}[1]{\langle {#1} \rangle}

\def\pb{\bar\psi}

\def\roughly#1{\mathrel{\raise.3ex\hbox{$#1$\kern-.75em%
\lower1ex\hbox{$\sim$}}}}
\def\lsim{\roughly<}
\def\gsim{\roughly>}
\def\psl{p\hspace{-1.7mm}/}
\def\qsl{q\hspace{-1.7mm}/}
\def\dfp{\frac{d^4 p}{(2\pi)^4}}

\def\={\;=\;}
\def\+{\;+\;}
\begin{document}
%
%\begin{titlepage}
%\pagestyle{empty}
%%\vspace{1.0in}
%\begin{flushright}
%preprint \#
%\end{flushright}
\begin{flushright}
 January 2000
\end{flushright}
\vspace{1.0cm}
\begin{center}
\doublespace
\begin{large}
{\bf Pion Properties in the $1/N_c$-corrected NJL model}\\
\end{large}
\vskip 1.0in
M. Oertel\footnote{e-mail:micaela.oertel@physik.tu-darmstadt.de}, M. Buballa
and J. Wambach\\
{\small{\it Institut f\"ur Kernphysik, TU Darmstadt,\\ 
Schlossgartenstr. 9, 64289 Darmstadt, Germany}}\\
\end{center}
\vspace{1cm}

\begin{abstract}
We investigate the effect of mesonic fluctuations on the pion propagator
in the Nambu--Jona-Lasinio (NJL) model by explicitly taking into
account $1/N_c$-corrections.
Because of the non-renormalizability of 
the model we have to regularize the meson loops with an independent 
cutoff parameter $\Lambda_M$. Whereas for moderate values of $\Lambda_M$
the pion properties change only quantitatively we encounter
strong instabilities for larger values of $\Lambda_M$.

\end{abstract}
%\end{titlepage}

%\newpage

\singlespace
%\doublespace

\section{Introduction}

It is generally believed that chiral symmetry, which is an approximate
symmetry of quantum chromodynamics, is spontaneously broken in vacuum.
Together with the linear sigma model the Nambu--Jona-Lasinio (NJL) model 
plays the role of a prototype for this mechanism. The original papers 
were published long time before QCD was developed \cite{NJL}.
By the observation that massless pions emerge as a consequence
of the spontaneous symmetry breaking they had direct impact on the 
formulation of the Goldstone theorem \cite{Gold}.
Today the NJL-model, which was originally a model of interacting
nucleons, is reinterpreted as a schematic quark model 
\cite{Kleinert,NJLrev}. 
Being much simpler than QCD, it is an attractive tool for studying
consequences of chiral symmetry breaking as well as its restoration
at large temperatures or densities, although its applicability 
to real processes is limited by several shortcomings of the model, in 
particular the lack of confinement and non-renormalizability. 

So far, most NJL model calculations have been performed in
leading order in $1/N_c$, the inverse number of colors.
Some authors have also considered corrections in next-to-leading order in
$1/N_c$, e.g. to
discuss $\pi-\pi$ scattering~\cite{Heidberg} or meson loop
corrections to the pion electromagnetic form factor~\cite{LT}. 
With the appropriate choice of model parameters chiral symmetry is
spontaneously broken in leading order, resulting in the 
appearance of massless pions and a massive sigma meson. 
If care is taken in the choice of correction terms~\cite{DSTL,Niko}, 
the pion emerges massless also in next-to-leading order. However, 
in a recent paper Kleinert and Van den Bossche argue that 
for the physical value, $N_c=3$, 
the spontaneous breakdown of chiral symmetry does not occur, as a
consequence of strong fluctuations, which come about as higher-order 
corrections in the $1/N_c$-expansion \cite{KVdB}. 
At present this paper is subject to many controversial discussions.

In the present article we want to study the effect of mesonic fluctuations 
on the pion propagator by calculating $1/N_c$-corrections explicitly. 
However, instead of trying to give a definite answer to the question
whether or not chiral symmetry is spontaneously broken,
we take the point of view that there {\it is} no unique solution in a
non-renormalizable model. At each order one encounters new divergences, 
which are in principle related to additional parameters (e.g. counter
terms). In the present case it is rather natural to introduce 
a new cutoff parameter $\Lambda_M$ to regularize the meson loops 
\cite{DSTL,Niko}.   
By varying this cutoff from zero to larger values we can
smoothly turn on the mesonic fluctuations and study their effects.

Although we cannot directly investigate the question of a possible
restoration of chiral symmetry within our scheme,
one might find hints for an instability.
This is a similar situation as in many-body systems,
where e.g. complex energy eigenvalues in the excitation spectrum 
indicate an instability of the ground state \cite{Thouless}.  
In fact, as we will discuss in Sec.~3, at large values of $\Lambda_M$
the pion propagator receives ``unphysical'' poles, i.e. poles at complex or 
space-like momenta. We also find negative wave function renormalization 
constants corresponding to residues with the wrong sign.
The main point we want to make in this letter is that the existence
of such signals depends on the choice of the meson
cutoff $\Lambda_M$. Obviously, for small values of $\Lambda_M$ the 
$1/N_c$-corrected results are close to the leading order ones
and no instability shows up. 
Ultimately, of course, the cutoff should be fixed by a fit to 
observables, for instance the decay width of vector mesons \cite{OBW}.

\section{The Model}

In this section we give a brief outline of our scheme for
describing mesons within the NJL model in next-to-leading order 
in $1/N_c$. To a large extent our model is based on Ref.~\cite{DSTL}.
More details will be presented elsewhere \cite{OBW}.

We consider the standard NJL Lagrangian for two flavors and three
colors with scalar-isoscalar and pseudoscalar-isovector interaction:
\be
   {\cal L} \;=\; \pb ( i \partial{\hskip-2.0mm}/ - m_0) \psi
            \;+\; g\,[(\pb\psi)^2 + (\pb i\gamma_5{\vec\tau}\psi)^2]  \ .
\ee
Here $g$ is a dimensionful coupling constant of order $1/N_c$. 
In the limit $m_0=0$ the Lagrangian is symmetric under 
$SU(2)_L \times SU(2)_R$ transformations. 

\begin{figure}[t]
\parbox{16cm}{\begin{center}
              \epsfig{file=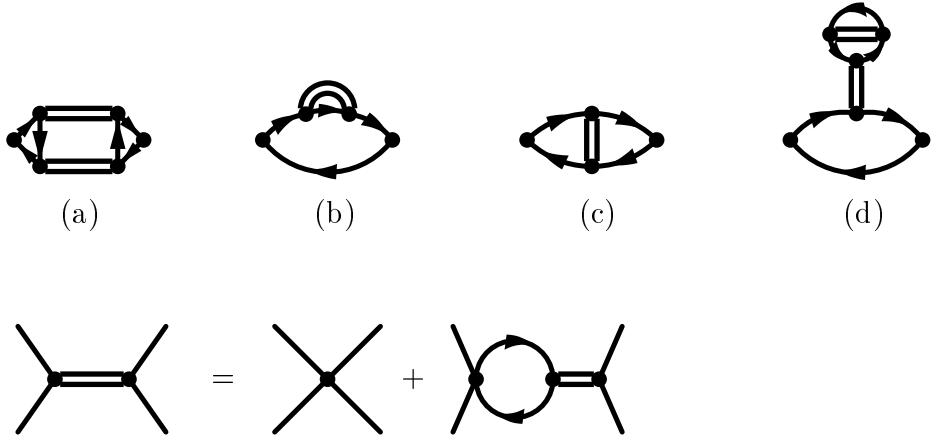}
              \end{center}}
\caption{\it Upper part: $1/N_c$-correction terms to the inverse meson 
propagators. The double lines denote meson propagators in leading order 
in $1/N_c$, which are calculated by iteration of quark-antiquark loops
as shown in the lower part of the figure.}
\label{fig1}
\end{figure}

In leading order in $1/N_c$ mesons are described by iterated 
quark-antiquark loops,
\be
   D_M^{q\bar q}(q^2) \= \frac{-2g}{1 \,-\, 2g\,\Pi_M^{q\bar q}(q^2)}  \;.
\label{DM}
\ee
This is illustrated in the lower part of  Fig.~\ref{fig1}.
For simplicity we will call $D_M^{q\bar q}$ a ``propagator'', although
strictly speaking it should be interpreted as 
$g_{Mqq}^2 \tilde D_M^{q\bar q}$, where $\tilde D_M^{q\bar q}$ is 
the renormalized meson propagator and $g_{Mqq}$ is a meson quark coupling 
constant. 
The quark-antiquark polarization diagrams are given by
\be
   \Pi_M^{q\bar q}(q^2) \= -i \, N_f N_c  \int \dfp \; tr[\,\Gamma_M \, 
   \frac{1}{\psl + \qsl - m + i\varepsilon} \, \Gamma_M \,
   \frac{1}{\psl - m + i\varepsilon}\,]              
   \;, 
\label{Piqq}
\ee
with $\Gamma_M = i\gamma_5$ for the pion and $\Gamma_M = 1$ for the
sigma channel. In Eq.~(\ref{Piqq})
$m$ denotes the constituent quark mass which is a solution of  
the gap equation
\be
   m \= m_0 + 2ig\; (N_f N_c) \int \dfp \; 
              tr \frac{1}{\psl - m +i\varepsilon}    \; . 
\label{gap}
\ee
It is now straight forward to show that for $m_0 = 0$ but $m\neq 0$ 
(``Nambu-Goldstone mode'') the right hand side of Eq.~(\ref{DM})
has a pole at $q^2 = 0$ in the pion channel and at $q^2 = 4m^2$ in the
sigma channel, corresponding to $m_\pi=0$ and $m_\sigma=2m$
\cite{NJL,NJLrev}. If $m_0/m$ is small but not exactly zero
the pion mass becomes non-zero and is approximately given by the
Gell-Mann Oakes Renner (GOR) relation \cite{GMOR}:
\be
   m_\pi^2 \,f_\pi^2 \= - m_0 \,\ave{\pb\psi}  \;.
\label{GOR}
\ee  

The $1/N_c$-correction terms to the inverse meson propagators, 
are shown in the upper part of Fig.~\ref{fig1}.
These diagrams contain mesonic fluctuations calculated in leading
order in $1/N_c$, as described above.
When iterated together with $\Pi_M^{q\bar q}$ the new terms lead to an 
improved meson propagator,
\be
   D_M(q^2) \= \frac{-2g}{1 \,-\, 2g\,(\Pi_M^{q\bar q}(q^2)
   + \delta\Pi_M(q^2) )}  \;,
\label{DMt}
\ee
where $\delta\Pi(q^2)$ denotes the sum of the correction terms.
It can be shown analytically that the pion constructed in this 
way is again massless in the chiral limit \cite{DSTL,OBW}.
Note that we do not use the improved meson propagators
$D_M(q^2)$ for evaluating the correction terms $\delta\Pi(q^2)$.
Such a selfconsistent procedure spoils the $1/N_c$ counting 
scheme and leads to inconsistencies with chiral symmetry. 

A non-renormalizable model is incomplete without defining how to regularize 
divergent loops. 
To leading order in $1/N_c$ one has to deal with quadratically divergent
integrals, the $q\bar q$ polarization diagrams (Eq.~(\ref{Piqq}))
and the gap equation (Eq.~(\ref{gap})).
These divergences can be regularized by standard methods, such as the
Pauli-Villars scheme or a sharp cutoff \cite{NJLrev}.
In next-to-leading order the situation is more involved. 
As an example we consider diagram (a) of Fig.~\ref{fig1}. 
This diagram contains two intermediate $q\bar q$-mesons, which have to be
calculated in a first step. We are then left with three loops: two
quark triangles and an integration over the four-momentum of the 
intermediate $q\bar q$-mesons. It is quite natural to regularize the
quark triangles in the same way as the $q\bar q$ polarization diagrams
which enter into the intermediate mesons. In fact, this is necessary
in order to preserve chiral symmetry. However, there is no stringent
reason, why the remaining meson loop should also be regularized in 
this way. We therefore follow Refs.~\cite{DSTL} and \cite{Niko} and
introduce a meson cutoff as an independent parameter. 

For computational convenience we work in the rest frame of the meson
(i.e. the external three-momentum is zero) and
regularize the internal meson loops by a 3-dimensional sharp cutoff 
$\Lambda_M$ in momentum space. On the other hand, since the internal 
mesons can of course not be chosen to be at rest, we should use a 
covariant scheme for their computation. As in Ref.~\cite{DSTL} we employ 
the Pauli-Villars scheme to regularize the quark loops. Because of 
the quadratic divergences we need two regulators. The regulator masses 
are chosen in the standard way as $M_1 = \sqrt{m^2 + \Lambda_q^2}$ and 
$M_2 = \sqrt{m^2 + 2\Lambda_q^2}$, with a free parameter $\Lambda_q$. 
This ``mixed'' scheme with a covariant regularization of the quark loops
and a non-covariant regularization of the meson loops was chosen for 
entirely practical reasons. However, we expect qualitatively similar 
results for other schemes, e.g. if we take a covariant sharp cutoff 
for both, quark and meson loops.

In this context we should also comment on Ref.~\cite{rostock} where
$1/N_c$-corrections to the quark condensate are studied within a NJL-type
model with a separable non-local interaction. This interaction generates
a form factor at the quark-vertices and can be chosen such that the 
3-momentum of each quark is limited to absolute values less than a
certain cutoff parameter. In this way also the 3-momentum of the meson
loops is automatically restricted without introducing an additional 
cutoff. Note that this does not contradict the statement that additional 
regulators are necessary if a non-renormalizable model is extended to higher 
loop orders: 
Because of the non-local interaction the model of Ref.~\cite{rostock}
is not identical to the NJL model but a 
modification which contains no longer any divergences. 
Although this is a quite appealing feature, the model has the disadvantage
of being manifestly non-covariant (the form factors depend on the 
3-momenta of the quarks). There are similar models with 4-momentum
dependent form factors (in Euclidean space) \cite{BK,BB} but these models 
have wrong analytical properties, which might be very disturbing in the
context of this article.    
Therefore - and because of the numerical effort - we decided to restrict
our investigations on the standard NJL model with the regularization scheme 
described above.

\section{Numerical Results}

We are now ready to study the influence of $1/N_c$-corrections on
the NJL pion propagator and related quantities.
We begin with the leading order, which corresponds to a meson cutoff 
$\Lambda_M$~=~0. With $\Lambda_q$~=~800~MeV, $g\Lambda_q^2$~=~2.9
and $m_0$~=~6.1 MeV we obtain a reasonable fit for the pion mass, 
the pion decay constant and the quark condensate:
$m_\pi^{(0)}$~=~140~MeV, $f_\pi^{(0)}$~=~93.5~MeV and
$\ave{\pb\psi}^{(0)}$~=~-2~(241.1~MeV)$^3$.
Here and in the following the superscript ${(0)}$ is used to denote
quantities which are calculated in leading order in $1/N_c$.
The above parameters correspond to a relatively small constituent
quark mass of 260~MeV.

\begin{figure}[t]
\parbox{16cm}{\begin{center}
              \epsfig{file=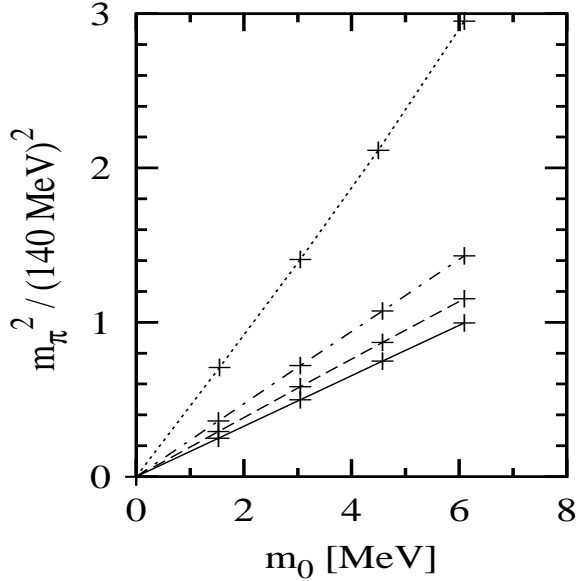,height=8cm,width=10cm}
              \end{center}}
\caption{\it Squared pion mass as a function of the current quark
             mass $m_0$ for different meson loop cutoffs:
             $\Lambda_M$~=~0~MeV (solid), 500~MeV (dashed), 
             900~MeV (dashed-dotted) and 1300~MeV (dotted). The calculated
             points are explicitly 
             marked.}
\label{fig2}
\end{figure}

Now we turn on the mesonic fluctuations by taking a non-zero 
meson cutoff $\Lambda_M$. Fig.~\ref{fig2} displays the squared pion
mass as a function of the current quark mass $m_0$ for different values
of $\Lambda_M$. Obviously all points which correspond to the same
meson cutoff lie almost on a straight line through the point
($m_0=0, m_\pi^2=0$). The latter was calculated analytically whereas
all other points are numerical results. 
This demonstrates the consistency of our scheme with chiral 
symmetry and the stability of the numerics.  

\begin{figure}[t] 
\parbox{16cm}{
  \hspace{-1.cm}
              \epsfig{file=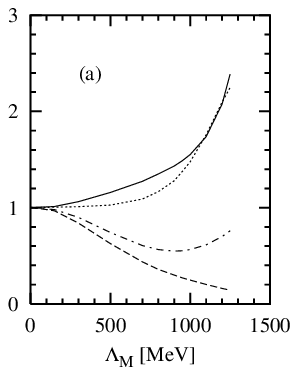,height=8cm,width=7cm}
              \hspace{2.cm}
              \epsfig{file=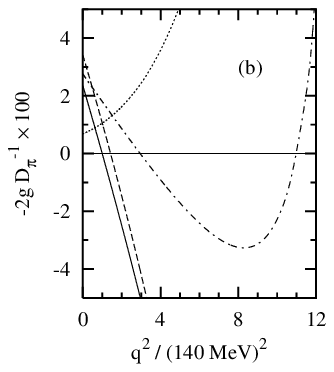,height=8cm,width=7cm}}
\caption{\it (a) The ratios $m_\pi^2/{m_\pi^2}^{(0)}$ (solid),
         $f_\pi^2/{f_\pi^2}^{(0)}$ (dashed), 
         $\ave{\pb\psi}/\ave{\pb\psi}^{(0)}$ (dashed-dotted)
         and the combination
         $-m_0\ave{\pb\psi}/m_{\pi}^2f_{\pi}^2$ (dotted)  
         as a function of the meson loop cutoff $\Lambda_M$. 
         (b) Inverse pion propagator, $D_\pi^{-1}$, multiplied by
         $-2g$ as a function of the 4-momentum squared for  various
         meson cutoffs: $\Lambda_M$~=~0 (solid), 900~MeV (dashed),
         1300~MeV (dashed-dotted) and 1500~MeV (dotted).  
          }
\label{fig3}
\end{figure}

Fig.~\ref{fig3}(a) displays the behavior of $m_\pi^2$, $f_\pi^2$
and the quark condensate with increasing $\Lambda_M$. All other parameters
are kept constant at the values given above. As one can see the mesonic
fluctuations lead to a reduction of $f_\pi$ while $m_\pi$ is increased. 
There are two $1/N_c$-correction terms to the quark condensate,  
which have been discussed, e.g., in Ref.~\cite{rostock}.
We find that the absolute value of the quark condensate decreases at smaller 
values of $\Lambda_M$ but goes up again for $\Lambda_M  \gsim$~900~MeV.

The leading-order quantities ${m_\pi^2}^{(0)}$, ${f_\pi^2}^{(0)}$ and 
$\ave{\pb\psi}^{(0)}$, are in almost perfect agreement with the GOR relation,
Eq.~(\ref{GOR}). From the (almost) linear dependence of $m_\pi^2$ on
$m_0$, as shown in Fig.~\ref{fig2}, one might expect that this is also
the case for $\Lambda_M > 0$. 
If one carefully expands both sides of Eq.~(\ref{GOR}) up to next-to-leading 
order in $1/N_c$, one can show analytically that our model is consistent with
the GOR relation \cite{OBW}. 
However, for the quantities 
$m_\pi^2$ and $f_\pi^2$ as they follow from the $1/N_c$-corrected inverse 
pion propagator, the l.h.s. of Eq.~(\ref{GOR}) also receives higher-order 
terms (i.e. terms beyond next-to-leading order)
in $1/N_c$, which are not present on the r.h.s. and therefore violate
the GOR relation. 
This can be seen in Fig.~\ref{fig3}(a), where the ratio of the 
right hand side and the left hand side of Eq.~(\ref{GOR}) is displayed
by the dotted line.
For $\Lambda_M\leq$~900~MeV the relation holds within 30\%.
However, when the meson cutoff is further increased the deviation
grows rapidly. This indicates that higher-order corrections in $1/N_c$
become important and our perturbative scheme should no longer be
trusted in this regime. 

The behavior of the pion mass becomes more clear from Fig.~\ref{fig3}(b) 
where the inverse pion propagator $D_\pi^{-1}(q^2)$ is plotted as a function 
of $q^2$ for different values of $\Lambda_M$. 
In the chiral limit all lines in this plot would go through 
$-2g D_\pi^{-1}$~=~0 at $q^2$~=~0. 
For $m_0 >$~0 the $q^2$~=~0-value of the leading-order result (solid line)
is shifted up to $m_0/m$, the ratio of current and constituent quark 
mass, and the curve crosses the zero-line at a positive value of $q^2$,
corresponding to a finite pion mass. 
For $\Lambda_M \lsim$~900~MeV the main effect of the $1/N_c$-corrections 
is a further increase of the $q^2$=0-value, while the slope does not
change significantly (dashed line). This leads to a moderate increase of the
pion mass. 
If we continue to increase $\Lambda_M$, the $q^2$=0-value of
$-2g D_\pi^{-1}$ moves down again. However, at the same time the function 
becomes more flat and this causes the accelerated increase of the pion mass
one sees in Fig.~\ref{fig3}(a). 
For $\Lambda_M \gsim$ 1250~MeV the curve even turns around at larger $q^2$ and 
crosses the zero line a second time (dashed-dotted line). 
At $\Lambda_M~\approx~$1350~MeV the two poles of $D_\pi$ merge and upon  
further increasing $\Lambda_M$ disappear from the positive real $q^2$-axis
(dotted line).

Obviously the pion becomes unstable if the mesonic fluctuations are too 
strong. In fact, already the second pole of the propagator in the regime
1250~MeV~$\lsim \Lambda_M \lsim$~1350~MeV is unphysical because the residue
has the 
wrong sign. This would correspond to an imaginary pion-quark coupling 
constant and to a negative value for $f_{\pi}^2$.
We found that this behavior is mainly due to diagram (b)  
in Fig.~\ref{fig1}. 
This diagram has a peak in the $q\bar q$ continuum which has an imaginary part
with the ``wrong'' sign. When $\Lambda_M$ exceeds a certain value, this
contribution becomes larger
than the sum of all other contributions, such that the imaginary part of
the inverse pion propagator,
$\mathrm{Im} D_{\pi}^{-1}$, gets the ``wrong'' sign. 
Via dispersion relations this peak can be related
to the turn-around of $\mathrm{Re} D_\pi^{-1}$ below the $q\bar q$ threshold
which is responsible for the second pole. Because of the unphysical features
of the second pole we used the first one to determine the pion mass for
$\Lambda_M$~=~1300~MeV in Fig.~\ref{fig2}.

It is quite reasonable that the instabilities of the pion propagator
indicate an instability of the underlying ground state with a spontaneously
broken chiral symmetry against mesonic fluctuations.
On the other hand it is clear that our model is not applicable to study
the process of chiral symmetry restoration itself.
Since $1/N_c$-corrections have been built in only perturbatively, 
the range of validity of the model is restricted to the regime where
these corrections are small. 
Therefore the rise of
$|\ave{\pb\psi}|$ for $\Lambda_M \geq$900~MeV should not be taken 
too serious.

The results of our calculations can be summarized by noting that
the structure of the pion propagator stays relatively stable for
$\Lambda_M \lsim \Lambda_q$ although sizable changes in 
$m_\pi$, $f_\pi$ and the quark condensate are found. For larger values of $\Lambda_M$ 
instabilities occur, which might be related to instabilities of the 
underlying ground state. 

Unfortunately, for numerical reasons we cannot perform calculations in the 
exact chiral limit, $m_0$~=~0, where the internal pion propagator
$D_\pi^{q\bar q}$ has a pole for $q^2=0$. However, as mentioned above, one can
show analytically for the improved pion propagator that $D_\pi^{-1}(0)$~=~0
in this case, independent of $\Lambda_M$. Furthermore, it is quite 
reasonable to expect that the slope of $D_\pi^{-1}$ will behave similarly
to what we found for $m_0$~=~6.1~MeV. This implies that above a 
certain value of $\Lambda_M$ the residue at $q^2$~=~0 gets again the 
``wrong'' sign, indicating an instability.

In Fig.~\ref{fig3} we did not change the parameters which were determined
in leading order by fitting $f_\pi^{(0)}$, $m_\pi^{(0)}$ and
$\ave{\pb\psi}^{(0)}$.
Of course, if one wants to apply the model to describe physical processes 
a refit of these observables should be performed including the 
$1/N_c$-corrections. This was partially done in Refs.~\cite{DSTL} and 
\cite{Niko} (without taking into account the full
momentum dependence of the quark triangles (see~Fig.~\ref{fig1} (a)) and the
other quark loops), 
where for each choice of $\Lambda_M$ the new $f_\pi$ was 
fitted to the empirical value. 
Since the $1/N_c$-corrections to $f_\pi$ 
are negative this means that the leading-order term $f_\pi^{(0)}$
has to be larger than the empirical value. As a consequence,
the onset of the instabilities is shifted to much larger values of 
$\Lambda_M$. In order to estimate the effect, we recall
that without refit the second pion pole shows up at
$\Lambda_M \simeq$~1250~MeV. At this point we find $f_\pi$~=~35~MeV
and $m_\pi$~=~216~MeV. A simple way to get the correct pion
decay constant $f_\pi^{exp}$~=~93~MeV is to rescale the parameters such 
that all quantities of dimension energy are multiplied by a factor 
$f_\pi^{exp}/f_\pi$, which is about 2.7 for the above example. 
Hence, the point at which the second pion pole emerges corresponds
to a rescaled cutoff $\Lambda_M \simeq$~3300~MeV. 
Of course, the pion mass is rescaled by the same factor and is now much 
too large ($\sim$~570~MeV). This can be easily cured by choosing 
a smaller current quark mass, which does not affect $f_\pi$ very much. 
Making use of the (almost) linear dependence of $m_\pi^2$ on $m_0$ we 
obtain $m_0 \simeq$~1~MeV.  
However, with these parameters we would strongly overestimate the quark 
condensate. Performing a more careful parameter study, it turns out
that a simultaneous fit of $f_\pi$ and $\ave{\pb\psi}$
is only possible for $\Lambda_M \lsim$~700~MeV.
Therefore the region where the instabilities show up in the pion
propagator seems to be far away from a realistic parameter set. 
Ultimately, one should of course try to determine also $\Lambda_M$ itself,
e.g. by fitting the decay width of vector mesons \cite{OBW}.

\section{Conclusions}

We have calculated the pion propagator within an NJL model which was 
extended to explicitly include meson loops in a $1/N_c$-expansion.
As already pointed out in Refs.~\cite{DSTL} and \cite{Niko}
this requires the introduction of a new cutoff parameter,
reflecting the non-renormalizability of the NJL model.
Similar to these authors an independent meson loop cutoff 
$\Lambda_M$ has been introduced in addition to the cutoff $\Lambda_q$ 
which was used to 
regularize the quark loops. Since the importance of the $1/N_c$-correction 
terms depends strongly on $\Lambda_M$, the question of how much 
the results are altered and in particular whether the spontaneous breakdown 
of chiral symmetry is spoiled by mesonic fluctuations \cite{KVdB}, cannot 
be uniquely answered.
Whereas for large values of $\Lambda_M$ instabilities show up in
the pion propagator, this is not the case at lower values of 
$\Lambda_M$. 
First estimates seem to indicate that the region of instabilities
is quite far away from a realistic parameter set, 
leaving enough room for further applications of the model to 
physical processes. 

The instabilities in the pion propagator might be a hint for instabilities 
of the underlying ground state against mesonic fluctuations. 
However, we should clearly state, that this question cannot be assessed 
within the formalism we have presented here.
Similarly the stable pion propagator we find at lower values
of $\Lambda_M$ does not prove that the ground state is stable.  
(Of course it should be stable if $\Lambda_M$ is sufficiently small.)
Here further careful studies are necessary including a comparison of 
different regularization schemes.

\section*{Acknowledgments}

We thank G. Ripka for illuminating discussions.
This work was supported in part by the BMBF and 
NSF grant NSF-PHY98-00978.

%\newpage
\singlespace

\end{document}